\documentclass[12pt]{iopart}
\usepackage{amssymb}

\begin{document}

\title{Intrinsic Conformal Symmetries in Szekeres models}
\author{Pantelis S. Apostolopoulos$^{1}$}

\address{$^1$Technological Educational Institute of Ionian Islands, Department
of Environmental Technology, Island of Zakynthos, Greece\\
Emails: papost@teiion.gr; papost@phys.uoa.gr}

\begin{abstract}
We show that Spatially Inhomogeneous (SI) and Irrotational dust models admit
a \emph{6-dimensional algebra }of \emph{Intrinsic Conformal Vector Fields}
(ICVFs) $\mathbf{X}_{\alpha }$ satisfying $p_{a}^{c}p_{b}^{d}\mathcal{L}_{%
\mathbf{X}_{\alpha }}p_{cd}=2\phi (\mathbf{X}_{\alpha })p_{ab}$ where $%
p_{ab} $ is the associated metric of the 2d distribution $\mathcal{X}$
normal to the fluid velocity $u^{a}$ and the radial unit spacelike vector
field $x^{a}$. The Intrinsic Conformal (IC) algebra is determined for each
of the curvature value $\epsilon $ that characterizes the structure of the
screen space $\mathcal{X}$. In addition the conformal flatness of the
hypersurfaces $\mathbf{u}=\mathbf{0}$ indicates the existence of a \emph{%
10-dimensional algebra} of ICVFs of the 3d metric $h_{ab}$. We illustrate
this expectation and propose a method to derive them by giving explicitly
the \emph{7 proper} ICVFs of the Lema\^{\i}tre-Tolman-Bondi (LTB) model
which represents the simplest subclass within the Szekeres family.
\end{abstract}

\maketitle

\noindent Szekeres spacetimes \cite{Szekeres:1974ct} can be regarded as being
quite general and, at the same time, \textquotedblleft
simple\textquotedblright\ enough within the vast set of Spatially
Inhomogeneous (SI) models. In the generic case they do not admit any kind of
global isometry \cite{Bonnor1, Bonnor-Sulaiman-Tomimura} nevertheless it has
been shown \cite{Krasinski} that represent a class of SI spacetimes with
special features \textquotedblleft restricting\textquotedblright , in some
sense, their generality. The key kinematical and dynamical aspects can be
collected as follows \cite{Apostolopoulos:2006zn, Apostolopoulos:2016xnm}%
\bigskip

\begin{enumerate}
\item The dust fluid velocity $u^{a}$ is geodesic and vorticity-free i.e. $%
\dot{u}_{a}=0=u_{[a;b]}$.

\item The magnetic part of the Weyl tensor vanishes identically i.e. $%
H_{ab}= $ $^{\bigstar }C_{acbd}u^{c}u^{d}=0$.

\item The Cotton-York tensor $C_{ab}$ of the $3-$dimensional hypersurfaces $%
t=$const. vanishes \cite{Krasinski} which implies their conformal flatness.

\item The symmetric tensor $h_{ab}=g_{ab}+u_{a}u_{b}$ is the proper metric
of the distribution $\mathbf{u}=\mathbf{0}$ with a well defined covariant
derivative $D_{a}$ that satisfies $D_{c}h_{ab}=h_{c}^{k}h_{a}^{i}h_{b}^{j}%
\nabla _{k}h_{ij}=0$.

\item There exist 3 independent spacelike unit vector fields $\left\{ 
\mathbf{x},\mathbf{y},\mathbf{z}\right\} $ that are hypersurface orthogonal $%
x_{[a}x_{b;c]}=y_{[a}y_{b;c]}=z_{[a}z_{b;c]}=0$. For each pair e.g. $\{%
\mathbf{u},\mathbf{x}\}$ the second order symmetric tensor $%
p_{ab}=h_{ab}-x_{a}x_{b}$ represents the proper metric of the screen space $%
\mathcal{X}$, which is defined as $\mathbf{u}\wedge \mathbf{x}=\mathbf{0}$,
with associated 2d derivative $\bar{\nabla}_{a}$. In particular $\bar{\nabla}%
_{c}p_{ab}=p_{c}^{k}p_{a}^{i}p_{b}^{j}\nabla _{k}p_{ij}=0$ and $%
\{u^{a},x^{a}\}$ are \emph{surface forming}: $p_{a}^{k}L_{\mathbf{u}}x_{k}=0$%
.

\item The 2d hypersurfaces of $t,r=$const. have constant curvature $\epsilon
=0,\pm 1$ \cite{Hellaby:2002nx}.\bigskip
\end{enumerate}

\noindent The coupling of the geometry with the dynamics, through the EFEs,
implies that there is a mutual influence between any geometrical or
dynamical constraint with subsequent restrictions in the structure of the
corresponding models. Therefore the inspection of properties (iii) and (vi)
allow us to examine the possibility of the existence of certain type of
symmetries inherent the Szekeres models. This is justified by two facts:
first it is well known \cite{Eisenhart} that every $n-$dimensional space of
constant curvature admits a $\frac{(n+1)\left( n+2\right) }{2}-$dimensional
algebra of Conformal Vector Fields (CVFs) with a $\frac{n\left( n+1\right) }{%
2}-$dimensional subalgebra of Killing Vector Fields (KVFs). Second a global $%
(1+n)-$decomposable space is conformally flat \emph{iff} the $n-$dimensional
space has constant curvature \cite{DecomposableConformal}, \cite{DecomposableConformalNDimensions}.

\noindent Although Szekeres models do not meet any of the above
requirements and do not admit any kind of symmetry, it follows from (iii) and (vi) that they are 
well fitted along the aforementioned lines. Motivated from these
arguements one should check for the existence of \emph{Intrinsic Conformal
Vector Fields} (ICVFs) \cite{Collins1, Collins-Szafron1, Collins-Szafron2, Collins-Szafron3} 
that generate transformations preserving the 
conformal structure of the 2d screen space $\mathcal{X}$ i.e. vector fields $%
\mathbf{X}$ satisfying 
\begin{equation}
p_{a}^{c}p_{b}^{d}\mathcal{L}_{\mathbf{X}}p_{cd}=\bar{\nabla}%
_{(b}X_{a)}=2\phi (\mathbf{X})p_{ab}  \label{DefinitionICVFs1}
\end{equation}%
\bigskip where $p_{a}^{k}X_{k}=X_{a}$ i.e. $\mathbf{X}$ are lying in the 2d
screen space $\mathcal{X}$.

\noindent Furthermore, the conformal flatness of the distribution $\mathbf{u}%
=\mathbf{0}$ together with the curvature ``constancy'' of $\mathcal{X}$
indicate that 
\begin{equation}
h_{a}^{c}h_{b}^{d}\mathcal{L}_{\mathbf{\Xi }}h_{cd}=D_{(b}\Xi _{a)}=2\phi (%
\mathbf{\Xi })h_{ab}.  \label{DefinitionICVFs2}
\end{equation}%
\bigskip for some 3-vector fields $\mathbf{\Xi }$ ($h_{a}^{k}\Xi _{k}=\Xi
_{a}$). However we postpone the detailed analysis and determination of the
10-dimensional algebra of ICVFs for a forthcoming work.

\noindent The purpose of this letter is to give the \emph{6-dimensional
algebra of ICVFs} admitted by the Szekeres models. Following \cite%
{Hellaby:2002nx} we rewrite the Szekeres metric in a form that manifests the
\textquotedblleft constancy\textquotedblright\ of the curvature of the 2d
screen space $\mathcal{X}$ 
\begin{equation}
ds^{2}=-dt^{2}+S^{2}\left\{ \frac{\left[ \left( \ln S/E\right) ^{\prime }%
\right] ^{2}}{\epsilon +F}dr^{2}+\frac{4\left( dy^{2}+dz^{2}\right) }{V^{2}%
\left[ \left( \frac{y-Y}{V}\right) ^{2}+\left( \frac{z-Z}{V}\right)
^{2}+\epsilon \right] ^{2}}\right\}
\label{SzekeresMetricHellabyKrasinskiForm1}
\end{equation}%
\bigskip where $S(t,r)$ represents the induced \textquotedblleft
scale\textquotedblright\ factor of the 2-dimensional screen space $\mathcal{X%
}$ \cite{Apostolopoulos:2016xnm}, $Y(r),$ $Z(r),$ $V(r),$ $F(r)$ are
arbitrary functions of their arguments and 
\begin{equation}
E(r,y,z)=\frac{V}{2}\left[ \left( \frac{y-Y}{V}\right) ^{2}+\left( \frac{z-Z%
}{V}\right) ^{2}+\epsilon \right]  \label{FunctionE1}
\end{equation}%
\bigskip controls the anisotropy of $\mathcal{X}$ and the way that
coordinates $(y,z)$ are mapped onto the unit sphere ($\epsilon =1$) , plane (%
$\epsilon =0$) or hyperboloid ($\epsilon =-1$) for each $r$.

\noindent Performing the change $V\hookrightarrow 2V$ the 2 dimensional line
element can be cast into the form 
\begin{equation}
ds_{2}^{2}=\frac{1}{V^{2}\left\{ \frac{1}{4V^{2}}\left[ \left( y-Y\right)
^{2}+\left( z-Z\right) ^{2}\right] +\epsilon \right\} ^{2}}\left(
dy^{2}+dz^{2}\right) .  \label{2dimensionalmetric1}
\end{equation}%
\bigskip

\noindent We consider two distinguished cases $\epsilon =\pm 1$ and $%
\epsilon =0$.\bigskip

\noindent \emph{Non vanishing 2d curvature (}$\epsilon =\pm 1$\emph{)}

\noindent From equation (\ref{2dimensionalmetric1}) it follows that the
metric of the hypersurface $\mathcal{X}$ (for $t,r=$const.) is written 
\begin{equation}
ds_{2}^{2}=\frac{d\left( y-Y\right) ^{2}+d\left( z-Z\right) ^{2}}{%
V^{2}\left\{ 1+\frac{k}{4}\left[ \left( y-Y\right) ^{2}+\left( z-Z\right)
^{2}\right] \right\} ^{2}}\equiv N^{2}ds_{\mathrm{FLAT}}^{2}
\label{NonZero2dCurvature1}
\end{equation}%
\bigskip where $k=\epsilon /V^{2}$ corresponds to the constant curvature%
\footnote{%
We recall that, locally, every $n-$dimensional space of constant curvature $%
k $ has a metric of the form $ds^{2}=\frac{dx^{i}dx_{i}}{\left( 1+\frac{k}{4}%
x_{i}x^{i}\right) ^{2}}$.} of the 2d space $\mathcal{X}$ for $t,r=$const.

\noindent Although the conformal algebra of a 2d space is infinite
dimensional we are interested only to a \emph{6-dimensional} subalgebra
consisting of the 3 independent KVFs and 3 \emph{gradient} CVFs. In
particular they can be found from the associated conformal algebra of the
flat metric $ds_{\mathrm{FLAT}}^{2}$ of the 2d space \cite%
{DecomposableConformal}, \cite{DecomposableConformalNDimensions} 
\begin{equation}
\mathbf{P}_{y}=\partial _{y},\quad \mathbf{P}_{z}=\partial _{z},\quad 
\mathbf{M}_{yz}=(z-Z)\partial _{y}-\left( y-Y\right) \partial _{z}
\label{KVFsFlat}
\end{equation}%
\begin{equation}
\mathbf{H}=\left( y-Y\right) \partial _{y}+\left( z-Z\right) \partial _{z}
\label{HVFFlat}
\end{equation}%
\begin{equation}
\mathbf{C}_{y}=2\left( y-Y\right) \mathbf{H}-\left[ \left( y-Y\right)
^{2}+\left( z-Z\right) ^{2}\right] \mathbf{P}_{y}  \label{SCVFyFlat}
\end{equation}%
\begin{equation}
\mathbf{C}_{z}=2\left( z-Z\right) \mathbf{H}-\left[ \left( y-Y\right)
^{2}+\left( z-Z\right) ^{2}\right] \mathbf{P}_{z}.  \label{SCVFzFlat}
\end{equation}%
\bigskip where the KVFs $\mathbf{P}_{y}$, $\mathbf{P}_{z}$, $\mathbf{M}_{yz}$
generate the 3d group of $y,z-$translations plus a rotation in the $yz-$%
plane, the Homothetic Vector Field (HVF) represents the scale-invariance
property of the flat space and $\mathbf{C}_{y}$, $\mathbf{C}_{z}$ are the
Special CVFs (SCVFs). 
\begin{equation}
\psi (\mathbf{P}_{y})=\psi (\mathbf{P}_{z})=\psi (\mathbf{M}_{yz})=0
\label{ConfFactors1}
\end{equation}%
\begin{equation}
\psi (\mathbf{H})=1,\quad \psi (\mathbf{C}_{y})=2\left( y-Y\right) ,\quad
\psi (\mathbf{C}_{z})=2\left( z-Z\right) .  \label{ConfFactors2}
\end{equation}%
\bigskip We note that $\left[ \psi (\mathbf{C}_{y})\right] _{,ab}=\left[
\psi (\mathbf{C}_{z})\right] _{,ab}=0$ which essentially reveal the complete
decomposability of (any dimension) flat Riemannian manifold. \bigskip

\noindent It follows that (\ref{KVFsFlat})-(\ref{SCVFzFlat}) are also CVFs
of the conformally related metric $ds_{2}^{2}$ and 
\begin{equation}
\phi (\mathbf{P}_{y})=-\frac{kN\left( y-Y\right) }{2},\quad \phi (\mathbf{P}%
_{z})=-\frac{kN\left( z-Z\right) }{2},\quad \phi (\mathbf{M}_{yz})=0
\label{ConfFactors3}
\end{equation}%
\begin{equation}
\phi (\mathbf{H})=1-\frac{k}{4}\left[ (y-Y)^{2}+(z-Z)^{2}\right] N
\label{ConfFactors4}
\end{equation}%
\begin{equation}
\phi (\mathbf{C}_{y})=2N(y-Y),\qquad \phi (\mathbf{C}_{z})=2N\left(
z-Z\right) .  \label{ConfFactors5}
\end{equation}%
\bigskip An appropriate linear combination of the conformal factors (\ref%
{ConfFactors3})-(\ref{ConfFactors5}) shows that the vectors $\mathbf{M}_{yz}$%
, $\frac{k}{4}\mathbf{C}_{y}+\mathbf{P}_{y}$, $\frac{k}{4}\mathbf{C}_{z}+%
\mathbf{P}_{z}$ are \emph{proper} KVFs and $\mathbf{H}$, $\frac{k}{4}\mathbf{%
C}_{y}-\mathbf{P}_{y}$, $\frac{k}{4}\mathbf{C}_{z}-\mathbf{P}_{z}$ are \emph{%
gradient} CVFs (i.e. their bivector vanishes indentically) of the constant
curvature metric $ds_{2}^{2}$. Collecting our results the conformal
subalgebra $\mathbf{X}_{\alpha }$ ($\alpha =1,..,6$) is 
\begin{equation}
\mathbf{X}_{1}=\mathbf{M}_{yz}  \label{FullSpaceKVF1}
\end{equation}%
\begin{equation}
\mathbf{X}_{2}=\left\{ 1+\frac{k}{4}\left[ \left( y-Y\right) ^{2}-\left(
z-Z\right) ^{2}\right] \right\} \partial _{y}+\frac{k}{2}\left( y-Y\right)
\left( z-Z\right) \partial _{z}  \label{FullSpaceKVF2}
\end{equation}%
\begin{equation}
\mathbf{X}_{3}=\frac{k}{2}\left( y-Y\right) \left( z-Z\right) \partial
_{y}+\left\{ 1+\frac{k}{4}\left[ \left( z-Z\right) ^{2}-\left( y-Y\right)
^{2}\right] \right\} \partial _{z}  \label{FullSpaceKVF3}
\end{equation}%
\begin{equation}
\mathbf{X}_{4}=\mathbf{H}=\left( y-Y\right) \partial _{y}+\left( z-Z\right)
\partial _{z}  \label{FullSpaceCVF1}
\end{equation}%
\begin{equation}
\mathbf{X}_{5}=\left\{ \frac{k}{4}\left[ \left( y-Y\right) ^{2}-\left(
z-Z\right) ^{2}\right] -1\right\} \partial _{y}+\frac{k}{2}\left( y-Y\right)
\left( z-Z\right) \partial _{z}  \label{FullSpaceCVF2}
\end{equation}%
\begin{equation}
\mathbf{X}_{6}=\frac{k}{2}\left( y-Y\right) \left( z-Z\right) \partial
_{y}+\left\{ \frac{k}{4}\left[ \left( z-Z\right) ^{2}-\left( y-Y\right) ^{2}%
\right] -1\right\} \partial _{z}.  \label{FullSpaceCVF3}
\end{equation}

\noindent \emph{Vanishing 2d curvature (}$\epsilon =0$\emph{)}

\noindent The flat metric (\ref{2dimensionalmetric1}) assumes the form 
\begin{equation}
ds_{2}^{2}=\frac{d\left( y-Y\right) ^{2}+d\left( z-Z\right) ^{2}}{\left\{ 
\frac{1}{4}\left[ \left( y-Y\right) ^{2}+\left( z-Z\right) ^{2}\right]
\right\} ^{2}}.  \label{Flat2Space1}
\end{equation}%
In the above, Riemann projection, coordinate system the KVFs are $\mathbf{X}%
_{1}=\mathbf{M}_{yz}$, $\mathbf{X}_{2}=\mathbf{C}_{y}$, $\mathbf{X}_{3}=%
\mathbf{C}_{z}$, the HVF is $\mathbf{X}_{4}=\mathbf{H}$ and the SCVFs are $%
\mathbf{X}_{5}=\mathbf{P}_{y}$, $\mathbf{X}_{6}=\mathbf{P}_{z}$ with 
\begin{equation}
\psi (\mathbf{C}_{y})=\psi (\mathbf{C}_{z})=\psi (\mathbf{M}_{yz})=0
\label{FlatSpaceConfFact1}
\end{equation}%
\begin{equation}
\psi (\mathbf{H})=-1,\quad \psi (\mathbf{P}_{y})=-2\frac{y-Y}{\left(
y-Y\right) ^{2}+\left( z-Z\right) ^{2}}  \label{FlatSpaceConfFact2}
\end{equation}%
\begin{equation}
\psi (\mathbf{P}_{z})=-2\frac{z-Z}{\left( y-Y\right) ^{2}+\left( z-Z\right)
^{2}}.  \label{FlatSpaceConfFact3}
\end{equation}%
\bigskip

\noindent Using the previous results, the lift to the Szekeres spacetime
shows that the vector fields (\ref{FullSpaceKVF1})-(\ref{FullSpaceCVF3}) ($%
\epsilon =\pm 1$) or (\ref{KVFsFlat})-(\ref{SCVFzFlat}) ($\epsilon =0$) are 
\emph{proper} ICVFs 
\begin{equation}
p_{a}^{c}p_{b}^{d}\mathcal{L}_{\mathbf{X}_{\alpha }}p_{cd}=\bar{\nabla}%
_{(b}X_{\alpha a)}=2\phi (\mathbf{X}_{\alpha })p_{ab}
\label{DefinitionICVFs3}
\end{equation}%
where $Y(r),$ $Z(r),$ $V(r)$ are functions of the radial coordinate and the
conformal factors are given by ($\epsilon =\pm 1$)%
\begin{equation}
\phi (\mathbf{X}_{1})=\phi (\mathbf{X}_{2})=\phi (\mathbf{X}_{3})=0
\label{ConfFactorsSzekeres1}
\end{equation}%
\begin{equation}
\phi (\mathbf{X}_{4})=\left\{ 1-\frac{k}{4}\left[ (y-Y)^{2}+(z-Z)^{2}\right]
\right\} N  \label{ConfFactorsSzekeres2}
\end{equation}%
\begin{equation}
\phi (\mathbf{X}_{5})=kN(y-Y),\qquad \phi (\mathbf{X}_{6})=kN\left(
z-Z\right) .  \label{ConfFactorsSzekeres3}
\end{equation}%
\bigskip and (\ref{FlatSpaceConfFact1})-(\ref{FlatSpaceConfFact3}) for the
case $\epsilon =0$.

\noindent We conclude this letter by noticing that the methodology presented
in \cite{DecomposableConformal}, \cite{DecomposableConformalNDimensions} could be helpful in order to get an
intuitive knowledge of how the \emph{10-dimensional algebra} of ICVFs of the 
$\mathbf{u}_{\perp }-$distribution of the Szekeres models is constituted. To
illustrate the tactic let us consider the Lema\^{\i}tre-Tolman-Bondi (LTB)
model \cite{tb}, \cite{bondi} as the simplest subclass of the Szekeres family 
\begin{equation}
ds^{2}=-dt^{2}+S^{2}\left\{ \frac{\left( S^{\prime }/S\right) ^{2}}{1+F}%
dr^{2}+\frac{\left( dy^{2}+dz^{2}\right) }{\left[ 1+\frac{1}{4}\left(
y^{2}+z^{2}\right) \right] ^{2}}\right\} .  \label{metrictb}
\end{equation}%
\bigskip We observe that the line element of the $t=$const. hypersurfaces is
conformally related with the \emph{1+2 decomposable} metric 
\begin{equation}
ds_{1+2}^{2}=d\tilde{r}^{2}+\frac{\left( dy^{2}+dz^{2}\right) }{\left[ 1+%
\frac{1}{4}\left( y^{2}+z^{2}\right) \right] ^{2}}
\label{1+2DecomposableLTB1}
\end{equation}%
\bigskip where we have used the gauge transformation $d\tilde{r}=\left[
\left( \ln S\right) ^{\prime }/\sqrt{1+F}\right] dr$ to redefine the radial
coordinate. The conformal algebra of the 1+2 decomposable metric is
described in terms of the 3 KVFs of the unit 2-sphere (which are expressed
in Cartesian coordinates), the KVF $\mathbf{\partial }_{\tilde{r}}=\left[
\left( \ln S\right) ^{\prime }/\sqrt{1+F}\right] ^{-1}\mathbf{\partial }_{r}$
and 6 CVFs (because for each gradient CVF of the 2d space we obtain 2 CVFs
of the $(1+2)-$manifold \cite{DecomposableConformal}, \cite{DecomposableConformalNDimensions}. It can be easily
verified that the vectors 
\begin{equation}
\mathbf{\hat{X}}_{5}=Ny\cosh \tilde{r}\partial _{\tilde{r}}+\sinh \tilde{r}%
\mathbf{X}_{5}  \label{Conformal1+2Vector1}
\end{equation}%
\begin{equation}
\mathbf{\hat{X}}_{6}=Ny\sinh \tilde{r}\partial _{\tilde{r}}+\cosh \tilde{r}%
\mathbf{X}_{5}  \label{Conformal1+2Vector2}
\end{equation}%
\begin{equation}
\mathbf{\hat{X}}_{7}=Nz\cosh \tilde{r}\partial _{\tilde{r}}+\sinh \tilde{r}%
\mathbf{X}_{6}  \label{Conformal1+2Vector3}
\end{equation}%
\begin{equation}
\mathbf{\hat{X}}_{8}=Nz\sinh \tilde{r}\partial _{\tilde{r}}+\cosh \tilde{r}%
\mathbf{X}_{6}  \label{Conformal1+2Vector4}
\end{equation}%
\begin{equation}
\mathbf{\hat{X}}_{9}=\phi (\mathbf{H})\cosh \tilde{r}\partial _{\tilde{r}%
}+\sinh \tilde{r}\mathbf{H}  \label{Conformal1+2Vector5}
\end{equation}%
\begin{equation}
\mathbf{\hat{X}}_{10}=\phi (\mathbf{H})\sinh \tilde{r}\partial _{\tilde{r}%
}+\cosh \tilde{r}\mathbf{H}  \label{Conformal1+2Vector6}
\end{equation}%
\bigskip are proper CVFs of (\ref{1+2DecomposableLTB1}). \bigskip

\noindent As we can check computationally, the vector fields $\mathbf{\hat{X}%
}_{4}=\left[ \left( \ln S\right) ^{\prime }/\sqrt{1+F}\right] ^{-1}\mathbf{%
\partial }_{r}$, $\mathbf{\hat{X}}_{5}$,...,$\mathbf{\hat{X}}_{10}$
correspond to the \emph{ICVFs} of the LTB model satisfying (\ref%
{DefinitionICVFs2}) and the \emph{arbitrary} function $\tilde{r}(t,r)$ is
determined, for each functional form of $S(t,r)$ and $F(r)$, by the equation 
\[
\frac{\partial \tilde{r}}{\partial r}=\frac{\left( \ln S\right) ^{\prime }}{%
\sqrt{1+F}}. 
\]%

\end{document}